\documentclass[journal=jacsat,manuscript=article]{achemso}
\usepackage[version=3]{mhchem}
\setkeys{acs}{articletitle = true}
\setkeys{acs}{maxauthors=10,etalmode=truncate}
\usepackage{chemformula} 
\usepackage[T1]{fontenc} 
\usepackage{enumitem}
\usepackage{graphicx}
\usepackage{titlecaps}
\Addlcwords{a an the of on at with for and in into from by without}

\usepackage{xr}
\newcommand*{\citen}[1]{%
  \begingroup
    \romannumeral-`\x 
    \setcitestyle{numbers}%
    \cite{#1}%
  \endgroup   
}

\usepackage{rotating}

\author{Debasish Koner}\affiliation[University of Basel]
       {Department of Chemistry, University of Basel,
         Klingelbergstrasse 80, 4056 Basel, Switzerland}

\author{Markus
  Meuwly}\email{m.meuwly@unibas.ch}\affiliation[University of Basel]
       {Department of Chemistry, University of Basel,
         Klingelbergstrasse 80, 4056 Basel, Switzerland}

\title[]{Permutationally Invariant, Reproducing Kernel-Based Potential
  Energy Surfaces for Polyatomic Molecules: From Formaldehyde to
  Acetone}

\begin{document}

\begin{abstract}
Constructing accurate, high dimensional molecular potential energy
surfaces (PESs) for polyatomic molecules is challenging.  Reproducing
Kernel Hilbert space (RKHS) interpolation is an efficient way to
construct such PESs. However, the scheme is most effective when the
input energies are available on a regular grid. Thus the number of
reference energies required can become very large even for
penta-atomic systems making such an approach computationally
prohibitive when using high-level electronic structure
calculations. Here an efficient and robust scheme is presented to
overcome these limitations and is applied to constructing high
dimensional PESs for systems with up to 10 atoms. Using energies as
well as gradients reduces the number of input data required and thus
keeps the number of coefficients at a manageable size. Correct
implementation of permutational symmetry in the kernel products is
tested and explicitly demonstrated for the highly symmetric CH$_4$
molecule.
\end{abstract}

\maketitle

\section{Introduction}
The dynamics of molecular system is entirely governed by the
underlying potential energy surface (PES) which describes the inter-
and intramolecular interactions. Often, such PESs are computed from
reference data based on electronic structure calculations using both,
regular or more random coordinate grids. As the study of the dynamics
of molecular systems requires energies and gradients, determining them
`{\it on the fly}' (i.e. {\it ab initio} molecular dynamics) can be
computationally prohibitive, in particular when high-level methods
such as second order M{\o}ller-Plesset (MP2), multi reference
configuration interaction (MRCI), or coupled cluster with singles,
doubles, and perturbative triples (CCSD(T)) are used together with
large basis sets. Therefore, constructing an analytical representation
of the ab initio PES is a meaningful and advantageous alternative to
accurately and efficiently describe intramolecular interactions.\\

\noindent
Developing accurate and computationally and data-efficient
representations of potential energies for multidimensional systems is
a challenging task. There are several approaches to describe the
energetics of a molecular PES: (i) fitting functional forms based on a
single or double many body expansion\cite{var07} such as the
London-Eyring-Polanyi-Sato (LEPS)\cite{por64:1105} or Aguado-Paniagua
(AP) surfaces,\cite{agu92:1265} (ii) permutationally invariant
polynomials (PIPs),\cite{qu18:151} (iii) interpolation by cubic
splines,\cite{xu05:244305} or modified Shepard
interpolation,\cite{she68:517,cre04:2392} (iv) kernel based methods
including reproducing kernel Hilbert space
(RKHS),\cite{ho96:2584,unk17:1923} Gaussian progress (GP)
regression,\cite{gp04} or (v) Neural network (NN) based
representations.\cite{beh07:146401,unk19:3678} The popular functional
terms (e.g. LEPS, AP) based on many body expansions can provide
accurate and computationally efficient representations for tri- and
tetra-atomic systems.\cite{kon13:13070,pau13:044309,kon16:034303}
However, using them for polyatomic systems is quite challenging as the
many body expansion becomes more complicated. Interpolation methods
are computationally expensive for multidimensional PESs whereas PIP,
GP, and NN approaches can be applied efficiently to construct
high-dimensional PESs.\cite{nan19:2826,unk19:3678,unk20:013001}\\

\noindent
RKHS interpolation has been shown to provide highly accurate PESs for
spectroscopic applications\cite{MM.n3:2019} and reaction
dynamics\cite{kon18:094305} as well as for molecular dynamics (MD)
simulations.  For small molecules (diatomic and
triatomic)\cite{hol01:3945,kon18:094305,kon19:24976,san20:3927,pez20:2171,kon20n2o}
this method is advantageous over other methods as it reproduces the
precalculated on-grid energies `exactly', captures the long range
interactions correctly if appropriate kernel polynomials are chosen
and results in smooth PESs with continuous
gradients.\cite{sol00:4415,ho00:3960} For a single energy evaluation
for an unknown molecular structure the RKHS method needs to sum over
all training samples.\cite{ho96:2584} However, if the {\it ab initio}
energies for training structures are provided on a regular grid, the
kernel functions can be decomposed into only two to five terms which
is much smaller than the training set size.\cite{hol97:7223} The sum
then runs over these few terms which can be precomputed and stored in
a look up table. Hence, with this {\it fast} RKHS approach the
computational cost scales almost linearly with the number of data
points\cite{hol97:7223,unk17:1923} and very accurate PESs can be
constructed for systems using a dense grid. The fast-evaluation method
was later modified to use partially filled grids with similar
efficiency.\cite{hol01:3940}\\

\noindent
It has been shown that within a high dimensional model representation
(HDMR), RKHS can be used to construct PESs. RKHS-HDMR works beyond
conventional tensor-product constructs and with successive multilevel
decomposition procedures which reduces multidimensional interpolation
to independent low dimensional interpolation.\cite{ho03:6433} This
approach can also be used for non-rectangular grids. An application of
the RKHS-HDMR approach to a low-dimensional (3d) system has been
reported for CH$_2$ as an example.\cite{ho03:6433} In a more recent
study, the RKHS-HDMR approach has been tested for the ten dimensional
Friedman target function but not for a PES.\cite{luo14:3099} However,
the use of RKHS for all degrees of freedoms in constructing PESs for
larger (i.e. four or more atoms) molecular systems is scarce in the
literature. Rather, a RKHS representation is used for selected degrees
of freedom, e.g. the van der Waals separation ($R$) whereas analytical
expressions are employed for the remaining degrees of freedom as was
done for tetra- and penta-atomic van der Waals complexes/molecules
e.g., OH--HCl\cite{wor05:244325}, HCN-HCl\cite{avo06:204315} and
NH$_3$--He.\cite{gub12:074301} \\
 
\noindent
One of the main difficulties in using grid-based interpolation methods
is their unfavourable scaling with increasing dimensionality of the
problem. Although the fast RKHS approach\cite{hol97:7223} allows for
near-independent data set size construction and evaluation of a RKHS,
the requirement of a rectangular grid-based reference data set
structure makes this approach highly computationally expensive in
terms of storage memory and number of operations. Even with a
partially filled grid the fast RKHS implementation scales as $2^M$
where $M$ is the number of dimensions/degrees of freedom, which makes
it unmanageable for more than four atom species. Sampling the
configuration space more densely near the stationary structures,
e.g. around minima and saddle points, can significantly reduce the
number of input energies.\cite{hol01:3940} But in practice using only
a small number of structures and energies leads to uneven RKHS PESs
with discontinuous gradients. On the other hand, including gradient
information for a configuration provides information about the likely
behavior of the PES in surrounding regions which is encoded in the
coefficients or parameters of an analytical PES. Hence, the analytical
PES provides a smooth behavior in the neighbourhood of a training grid
point even if only fewer numbers of configurations are used for
training.\\

\noindent
It has been shown for permutationally invariant polynomials (PIPs)
applied to CH$_4$ that by using gradients along with energies in the
input data set, smooth and accurate PESs can be obtained using fewer
input data.\cite{nan19:2826} From energy and gradient information for
only 100 configurations, randomly selected from an {\it ab initio}
molecular dynamics (AIMD) simulation, a PIP-based PES was constructed
with root mean square errors of 8.8 cm$^{-1}$ and 39.8 cm$^{-1}$/a$_0$
for energy and gradients, respectively. The harmonic frequencies from
the normal mode analysis using those PIP PESs were within 1 cm$^{-1}$
compared with the {\it ab initio} frequencies. Subsequently, this
approach was applied to N-methyl acetamide (NMA) to construct PESs for
{\it trans}-NMA\cite{qu19:141101} and a full dimensional PES for
NMA\cite{nan19:084306} with a root mean squared fitting error ranging
from 26.8 cm$^{-1}$ for full PIP and 148.9 cm$^{-1}$ when a
fragment-based approach was used whereby the energies used in the
fitting covered a range up to $\sim 3.5$ eV.\\
 
\noindent
Here, we introduce an efficient and robust approach to represent
highly accurate PESs for molecules with four to ten atoms using RKHS
interpolation with reciprocal power decay kernels. Gradients are used
along with the energies to determine the coefficients for the tensor
product form of the kernels. The formulation is applied to systems
ranging from formaldehyde (CH$_2$O, 4 atoms) to acetone
(CH$_3$COCH$_3$, 10 atoms). Molecular symmetry is included explicitly
in the tensor product expansion of the kernel polynomials and is
demonstrated to yield accurate RKHS-based results for the highly
symmetric CH$_4$ molecule. First, the methodological developments are
discussed. Next, RKHS-based PESs are determined for illustrative
examples and the harmonic frequencies are determined as a validation
of the methods. Finally, conclusions are drawn.\\
 
\section{Methods}
\subsection{RKHS with Energies and Gradients}
Within the RKHS formalism\cite{aronszajn1950rkhs} potential energies
for a system can be expressed as a linear combination of reproducing
kernel functions using a set of known energies $ V({\bf x})$ at
different configurations ${\bf x}$. The representer
theorem\cite{scholkopf2001generalized} for a general functional
relationship $y = f(\mathbf{x})$ states that $f(\mathbf{x})$ can
always be approximated as a linear combination of suitable functions
\begin{equation}
f(\mathbf{x}) \approx \widetilde{f}(\mathbf{x}) = \sum_{i = 1}^{N}
\alpha_i K(\mathbf{x},\mathbf{x}_i)
\label{eq:kernel_regression}
\end{equation}
where $\alpha_i$ are coefficients and $K(\mathbf{x},\mathbf{x'})$ is a
kernel function. The reproducing property asserts that $f(x') =
\langle f(x),K(x,x') \rangle$ where $\langle \cdot \rangle$ is the
scalar product and $K(x,x')$ is the kernel.\cite{aronszajn1950rkhs}
Popular choices for $K(\mathbf{x},\mathbf{x'})$ for representing PESs
are polynomial kernels
\begin{equation}
K(\mathbf{x},\mathbf{x'}) = \langle\mathbf{x},\mathbf{x'}\rangle^d
\label{eq:polynomial_kernel}
\end{equation}
where $\langle \cdot, \cdot \rangle$ denotes the dot product and $d$
is the degree of the polynomial. It is also possible to include
knowledge about the long range behaviour of the physical interactions
into the kernel function
itself.\cite{hollebeek.annrevphychem.1999.rkhs,sol00:4415}\\

\noindent
The coefficients $\alpha_i$ (Eq.~\ref{eq:kernel_regression}) can be
determined such that $\widetilde{f}(\mathbf{x}_i) = y_i$ for all input
$\mathbf{x}_i$ in the dataset, i.e.\
\begin{equation}
	\boldsymbol{\alpha} = \mathbf{K}^{-1}\mathbf{y}
	\label{eq:krr_coefficient_relation}
\end{equation}
where $\boldsymbol{\alpha} = \left[\alpha_i \cdots
  \alpha_N\right]^\mathrm{T}$ is the vector of coefficients,
$\mathbf{K}$ is an $N\times N$ matrix with entries $K_{ij} =
K(\mathbf{x}_i,\mathbf{x}_j)$ called kernel
matrix\cite{muller2001introduction,hofmann2008kernel} and $\mathbf{y}
= \left[y_1 \cdots y_N\right]^\mathrm{T}$ is a vector containing the
$N$ observations $y_i$ in the data set. Since the kernel matrix is
symmetric and positive-definite by construction, Cholesky
decomposition\cite{golub2012matrix} can be used to efficiently solve
Eq.~\ref{eq:krr_coefficient_relation}.  Once the coefficients
$\alpha_i$ have been determined, unknown values $y_*$ at arbitrary
positions $\mathbf{x}_*$ can be estimated as
$y_*=\widetilde{f}(\mathbf{x}_*)$ using
Eq.~\ref{eq:kernel_regression}. \\

\noindent
In practice the solution of Eq.~\ref{eq:krr_coefficient_relation} is
only possible if the kernel matrix $\mathbf{K}$ is not
ill-conditioned. Fortunately, even if $\mathbf{K}$ is ill-conditioned,
an approximate (regularized) solution can be obtained for example by
Tikhonov regularization\cite{tikhonov1977solutions}. This amounts to
adding a small positive constant $\lambda$ to the diagonal of
$\mathbf{K}$, such that
\begin{equation}
    \boldsymbol{\alpha} = \left(\mathbf{K}+\lambda\mathbf{I}\right)^{-1}\mathbf{y}
	\label{eq:krr_coefficient_relation_regularized}
\end{equation}
is solved instead of Eq. \ref{eq:krr_coefficient_relation} when
determining the coefficients $\alpha_i$ (here, $\mathbf{I}$ is the
identity matrix). Adding $\lambda > 0$ to the diagonal of $\mathbf{K}$
damps the magnitude of the coefficients $\boldsymbol{\alpha}$ and
increases the smoothness of $\widetilde{f}$. While this has the effect
that the known values in the data set are only {\it approximately}
reproduced by Eq.~\ref{eq:kernel_regression}, i.e.\ strictly
$\widetilde{f}(\mathbf{x}_i) \neq y_i$, perhaps counterintuitively, it
can {\it increase} the overall quality of predictions for unknown
$\mathbf{x_*}$: In cases where the values $y_i$ are noisy, reproducing
them exactly also reproduces the noise, which is unlikely to
generalise well to unknown data. Therefore, this method of determining
the coefficients can also be used to prevent over-fitting and is known
as kernel ridge regression (KRR).\\

\noindent
When applied to represent discrete data for energies, the PES can be
written as
\begin{equation}
     V({\bf x}) = \sum_{i=1}^N \alpha_i K({\bf x}, {\bf x'_i})
\end{equation}
\label{eq:eq1}
where $\alpha_i$ are coefficients and $K({\bf x}, {\bf x'})$ is the
reproducing kernel and ${\bf x_i'}$ represents the training set which
are the geometries for which energies have been determined from
electronic structure calculations. The coefficients are then
determined from the known ab initio energies for $N$ configurations by
solving the linear equations
 \begin{equation}
\begin{pmatrix} K ({\bf{x}}_1, {\bf{x'}}_1) & K ({\bf{x}}_1, {\bf{x'}}_2) & \cdots & K ({\bf{x}}_1, {\bf{x'}}_N) \\
K ({\bf{x}}_2, {\bf{x'}}_1) & K ({\bf{x}}_2, {\bf{x'}}_2) & \cdots & K ({\bf{x}}_2, {\bf{x'}}_N) \\
\vdots & \vdots & \ddots & \vdots \\
K ({\bf{x}}_N, {\bf{x'}}_1) & K ({\bf{x}}_N, {\bf{x'}}_2) & \cdots & K ({\bf{x}}_N, {\bf{x'}}_N) \\
\end{pmatrix} \begin{pmatrix}
 \alpha_1 \\
 \alpha_2 \\
 \vdots \\
 \alpha_N \\
\end{pmatrix}=\begin{pmatrix}
 V_1 \\
 V_2 \\
 \vdots \\
 V_N \\
\end{pmatrix}
\label{rkhs}
\end{equation}
This procedure gives an exact solution on the grid points ${\bf
  x_i'}$. The explicit matrix form (Eq. \ref{rkhs}) for
Eq. \ref{eq:kernel_regression} is given to clarify how the structure
of $K(\bf{x},\bf{x_i})$ changes once gradients are included in
constructing the RKHS (see below).\\
  
\noindent
For an $M$-dimensional problem, the multi-dimensional kernel can be
written as a direct product
\begin{equation}
   K({\bf x}, {\bf x'}) = \prod_{j=1}^M k_j(x, x')
\label{mdker}
\end{equation}
where $k_j(x, x')$ are 1D kernels. Multidimensional reproducing
kernels can therefore be used to represent the $p$-body interaction
energies of a system.\\

\noindent
Within a many body expansion, the total potential energy of a system
can be decomposed into a sum of $p$-body interactions $V^{(p)}$. For a
molecule with $n$ atoms, each $p$-body term consists of $^nC_p$
$p$-body interactions, where $^nC_p$ is the binomial coefficient. The
total potential for an $n$-atomic species is therefore
 \begin{equation}
  V = \sum_{p=1}^n \sum_{i=1}^{^nC_p} V_i^{(p)}
\label{mbe}
 \end{equation}
In practice Eq. \ref{mbe} is truncated at $p=3$ or 4,
i.e. contributions up to three- and 4-body terms are included which is
what is also done in the present work.\\
 
\noindent
One dimensional, reciprocal power reproducing kernels have been shown
to describe diatomic potentials with high accuracy on the interval
$[0,\infty]$.\cite{ho96:2584,sol00:4415} The general expression for a
$k^{[n,m]}$ reproducing polynomial kernel is
 \begin{equation}
 k^{[n,m]} = n^2x_{>}^{-(m+1)}B(m+1,n)
 _2F_1\left(-n+1,m+1;n+m+1;\frac{x_<}{x_>}\right )
\label{rprk}
\end{equation}
where, $n$ and $m$ are the smoothness and asymptotic reciprocal power
parameters, whereas $x_<$ and $x_>$ are the smaller and larger value
of $x$, respectively. $B(a,b)$ in Eq. \ref{rprk} is the beta function
$B(a,b) = \frac{(a-1)!(b-1)!}{(a+b-1)!}$ and $_2F_1(a,b;c;z)$ is
Gauss' hypergeometric function.\cite{ho96:2584} These kernel
polynomials can also be used to construct an $M$-dimensional
reproducing kernels as a function of radial dimensions by direct
product relations. In the present study, each term of $p$-body
interaction energy is represented as an $M$-dimensional ($M = $
$^pC_2$) reproducing kernel constructed from $M$ reciprocal power
kernels for $M$ interatomic distances $r_j$. The full kernel is then
\begin{equation}
K({\bf r}, {\bf r'}) = \sum_{p=1}^n
\sum_{l=1}^{^nC_p}\prod_{j=1}^{^pC_2} k_j(r_j,r'_j)
\label{fullkernel1}
 \end{equation}
and
  \begin{equation}
     V({\bf r}) = \sum_{i=1}^N \alpha_i K({\bf r}, {\bf r'})
     \label{fullkernel2}
 \end{equation}
Here, ${\bf r}$ is a vector containing all pairwise interatomic
distances of an $n$-atomic system, ${\bf r} = \{r_h|h=1,2,3
\cdots,^nC_2$\}.  In this study different reciprocal power kernels
were tested, and it is found that $k^{[3,5]}$, $k^{[3,1]}$ and
$k^{[3,0]}$ kernels perform best to construct mono/multidimensional
kernels for 2-, 3-, and 4-body interaction energies, respectively.\\
 
\noindent
Derivatives of the potential with respect to the distance coordinates
can be calculated by simply replacing the reproducing kernels $K({\bf
  r}, {\bf r'})$ by their derivatives $K'({\bf r}, {\bf r'})$. Then
the gradients of the total potential with respect to a Cartesian
coordinates $x_i$ are
\begin{equation}
  \frac{dV}{dx_i} = \sum_{h=1}^{^nC_2}\frac{dV}{dr_h}\frac{dr_h}{dx_i}
  \label{derkernel1}
\end{equation}
and
\begin{equation}
\frac{dV}{dr_h}=\sum_{i=1}^N C_i K'({\bf r}, {\bf r'})
  \label{derkernel2}
\end{equation}
If the PES is faithfully represented by the RKHS, its derivative is
also a good approximation of the gradients.\\

\noindent
In a next step, the gradients - which are also available from the
electronic structure calculations - are included in Eq. \ref{rkhs}
which yields
\begin{equation}
\begin{pmatrix} K ({\bf{x}}_1, {\bf{x'}}_1) & K ({\bf{x}}_1, {\bf{x'}}_2) & \cdots & K ({\bf{x}}_1, {\bf{x'}}_N) \\
K'_{x1} ({\bf{x}}_1, {\bf{x'}}_1) & K'_{x1} ({\bf{x}}_1, {\bf{x'}}_2) & \cdots & K'_{x1}({\bf{x}}_1, {\bf{x'}}_N) \\
 K'_{y1} ({\bf{x}}_1, {\bf{x'}}_1) & K'_{y1} ({\bf{x}}_1, {\bf{x'}}_2) & \cdots & K'_{y1}({\bf{x}}_1, {\bf{x'}}_N) \\
K'_{z1} ({\bf{x}}_1, {\bf{x'}}_1) & K'_{z1} ({\bf{x}}_1, {\bf{x'}}_2) & \cdots & K'_{z1}({\bf{x}}_1, {\bf{x'}}_N) \\
 \vdots & \vdots & \ddots & \vdots \\
 K'_{xn} ({\bf{x}}_1, {\bf{x'}}_1) & K'_{xn} ({\bf{x}}_1, {\bf{x'}}_2) & \cdots & K'_{xn}({\bf{x}}_1, {\bf{x'}}_N) \\
 K'_{yn} ({\bf{x}}_1, {\bf{x'}}_1) & K'_{yn} ({\bf{x}}_1, {\bf{x'}}_2) & \cdots & K'_{yn}({\bf{x}}_1, {\bf{x'}}_N) \\
K'_{zn} ({\bf{x}}_1, {\bf{x'}}_1) & K'_{zn} ({\bf{x}}_1, {\bf{x'}}_2) & \cdots & K'_{zn}({\bf{x}}_1, {\bf{x'}}_N) \\
 \vdots & \vdots & \ddots & \vdots \\
 
 K ({\bf{x}}_N, {\bf{x'}}_1) & K ({\bf{x}}_N, {\bf{x'}}_2) & \cdots & K ({\bf{x}}_N, {\bf{x'}}_N) \\
 K'_x ({\bf{x}}_N, {\bf{x'}}_1) & K'_x ({\bf{x}}_N, {\bf{x'}}_2) & \cdots & K'_x({\bf{x}}_N, {\bf{x'}}_N) \\
 K'_y ({\bf{x}}_N, {\bf{x'}}_1) & K'_y ({\bf{x}}_N, {\bf{x'}}_2) & \cdots & K'_y({\bf{x}}_N, {\bf{x'}}_N) \\
K'_z ({\bf{x}}_N, {\bf{x'}}_1) & K'_z ({\bf{x}}_N, {\bf{x'}}_2) & \cdots & K'_z({\bf{x}}_N, {\bf{x'}}_N)\\
 \vdots & \vdots & \ddots & \vdots \\
 K'_{xn} ({\bf{x}}_N, {\bf{x'}}_1) & K'_{xn} ({\bf{x}}_N, {\bf{x'}}_2) & \cdots & K'_{xn}({\bf{x}}_N, {\bf{x'}}_N) \\
 K'_{yn} ({\bf{x}}_N, {\bf{x'}}_1) & K'_{yn} ({\bf{x}}_N, {\bf{x'}}_2) & \cdots & K'_{yn}({\bf{x}}_N, {\bf{x'}}_N) \\
K'_{zn} ({\bf{x}}_N, {\bf{x'}}_1) & K'_{zn} ({\bf{x}}_N, {\bf{x'}}_2) & \cdots & K'_{zn}({\bf{x}}_N, {\bf{x'}}_N) \\
\end{pmatrix}
 \begin{pmatrix}
  \alpha_1 \\
  \alpha_2 \\
  \vdots \\
  \alpha_N \\
  \end{pmatrix}
 =\begin{pmatrix}
  V_1 \\
  dV_1/dx1 \\
dV_1/dy1 \\
dV_1/dz1 \\
  \vdots \\
    dV_1/dxn \\
dV_1/dyn \\
dV_1/dzn \\
  \vdots \\
  V_N \\
   dV_N/dx1 \\
dV_N/dy1 \\
    dV_N/dz1 \\
      \vdots \\
   dV_N/dxn \\
dV_N/dyn \\
    dV_N/dzn \\
 \end{pmatrix}
 \label{rkhs+f}
\end{equation}
For a species with $n$ atoms and $N$ configurations ${\bf x}$ for
which energies have been computed, the left-hand side matrix in
Eq. \ref{rkhs+f} has dimension $(3n+1)N\times N$. Eq. \ref{rkhs+f} can
be solved using a least square fitting algorithm. Here, the `dgelss'
subroutine from the LAPACK library\cite{lapack} is used to solve the
set of linear equations.\\

\noindent
To better represent important (i.e. low-energy) regions of the PES, a
weighted fit is performed. The weights $w_i$ for each point have been
chosen as
\begin{equation}
    w_i=\frac{\Delta V}{\Delta V + (V_i-V_{\rm min})}
    \label{wenergy}
\end{equation}
where $\Delta V$ is either a constant (here 4 eV) or the maximum
energy of the training set relative to the minimum ($\Delta V = V_{\rm
  max}-V_{\rm min}$), and $V_i$ is the relative energy of a
configuration $i$ with respect to the minimum energy of the system
$V_{\rm min}$. In this way, a larger weight is assigned to structures
close to the equilibrium. A similar weight function is also used for
the gradients
\begin{equation}
    w_i=\frac{\Delta g}{\Delta g + |g_i|}.
     \label{wgradient}
\end{equation}
The maximum value of $\Delta g$ is 10 eV/$a_0$.\\

\subsection{Symmetrized RKHS}
One of the main challenges when constructing a multidimensional PES is
to maintain the symmetry of the PES with respect to interchanging
equivalent atoms. Configurations for all permutations of equivalent
atoms are to be included. The most straightforward way is to include
all permutationally equivalent configurations with the same energies
in the training data set. However, this increases the size of the
training data set, which also increases the evaluation cost in RKHS
for an energy evaluations the sum runs for all the training
structures. Also to obtain the coefficients the set of linear
equations are solved numerically which may lead to a mismatch between
energies of two equivalent structures due to numerical
inaccuracies. Hence, it is advantageous to rather explicitly
symmetrize the total kernel polynomial $K({\bf r}, {\bf r'})$ (see
Eq. \ref{fullkernel1}) by expanding it as a linear combination of all
equivalent structures of a molecule.
\begin{equation}
    K_{\rm sym}({\bf r}, {\bf r'}) =\sum_{i=1}^S K_{i}({\bf r}, {\bf r'}),
     \label{symkernel}
\end{equation}
where $S$ is the number of equivalent configurations. A similar
strategy was followed in constructing PESs from PIPs for which
symmetrized basis functions were generated by adding products of all
`monomials' for a molecule considering permutations of equivalent
atoms.\cite{bra09:577}\\

\begin{figure}[h!]
\centering \includegraphics[scale=1.2]{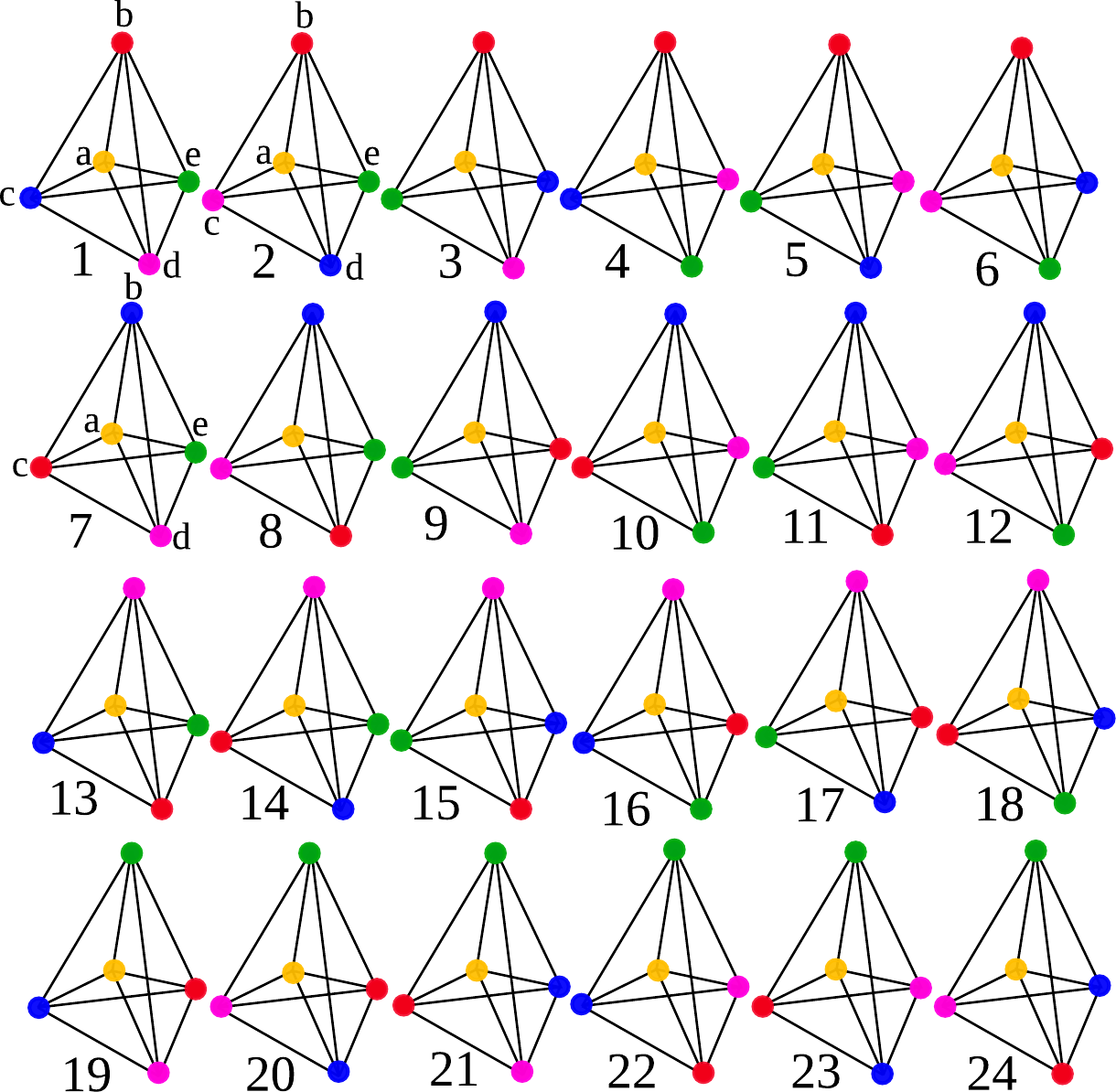}
\caption{All 24 permutations of H atoms in CH$_4$ molecule. Atoms are
  represented by color, yellow (y) is for the carbon atom and red (r),
  blue (b), magenta (m) and green (g) for the hydrogen
  atoms. Positions of the atoms are denoted by `a', `b', `c', `d' and
  `e'.}
\label{fig:fig1}
\end{figure}

\noindent
An example is given here for the CH$_4$ molecule. All permutations
with respect to four equivalent H atoms are shown in Figure
\ref{fig:fig1}. Atom positions are assigned by `a' through `e', while
different atoms can be distinguished by different colors.  The order
of the interatomic distances with respect to positions are given in
Table \ref{tab:sym} for all permutations. For CH$_4$ there are four
and six equivalent CH and HH distances, respectively. To define a 1D
kernel two bond distances are required: either $k(x,x')$ or $k(y,z')$
where $x$ and $x'$ are the same pairwise distance (here the C-H or H-H
distances) and $y$ and $z'$ are two distances that need to be
explicitly symmetrized (here two H-H or two C-H distances for
symmetry-related hydrogen atoms). In the absence of symmetry, ten 1D
kernels (($1^2 \times 4$) + ($1^2 \times 6$)) for interatomic
distances define the basis set for RKHS $k(r_{\rm ab},r_{\rm ab}'),
k(r_{\rm ac},r_{\rm ac}'), \cdots, k(r_{\rm de},r_{\rm de}')$ (only
one configuration is possible e.g. configuration 1 in Figure
\ref{fig:fig1}). However, using symmetry each configuration has 24
permutations which leads to 52 1D kernels ($4^2 + 6^2 = 52$ for the
four CH and six HH bonds) for interatomic distances to complete the
basis set for RKHS. All 52 1D basis kernel functions are reported in
Table \ref{tab:sym} i.e. $[k(r_{\rm yr},r_{\rm yr}'), \cdots, k(r_{\rm
    mg},r_{\rm mg}')]$, $[k(r_{\rm yr},r_{\rm yr}'), \cdots, k(r_{\rm
    mg},r_{\rm bg}')]$, $\cdots, [k(r_{\rm yr},r_{\rm yg}'), \cdots,
  k(r_{\rm mg},r_{\rm rb}')]$. It is to be noted that Table
\ref{tab:sym} contains 240 kernel functions in total whereas many of
them ($(6\times(4\times4)+4\times(6\times6))$ are equivalent. The
2-body interaction energy is then the sum of all these 240 1D kernel
functions.\\

\begin{table}[ht!]
\caption{Symmetrization order of interatomic distances for equivalent
  CH$_4$ structures. Interatomic distances between two different
  atoms/positions are $r_{ij}$ = $r_{ji}$. Atom positions and color
  indices are defined in Figure \ref{fig:fig1}.}
    \begin{tabular}{l|cccc|cccccc}
\hline
\hline
Configurations & $r_{\rm ab}$ & $r_{\rm ac}$ & $r_{\rm ad}$ & $r_{\rm ae}$ & $r_{\rm bc}$ & $r_{\rm bd}$ & $r_{\rm be}$ & $r_{\rm cd}$ & $r_{\rm ce}$ & $r_{\rm de}$  \\
\hline
1  & $r_{\rm yr}$ & $r_{\rm yb}$ & $r_{\rm ym}$ & $r_{\rm yg}$ & $r_{\rm rb}$ & $r_{\rm rm}$ & $r_{\rm rg}$ & $r_{\rm bm}$ & $r_{\rm bg}$ & $r_{\rm mg}$  \\

2 & $r_{\rm yr}$ & $r_{\rm ys}$ & $r_{\rm yb}$ & $r_{\rm yg}$ & $r_{\rm rm}$ & $r_{\rm rb}$ & $r_{\rm rg}$ & $r_{\rm mb}$ & $r_{\rm mg}$ & $r_{\rm bg}$  \\
$\vdots$ & $\vdots$ & $\vdots$ &$\vdots$ &$\vdots$ &$\vdots$ &$\vdots$ &$\vdots$ &$\vdots$ &$\vdots$ &$\vdots$ \\
23 & $r_{\rm yg}$ & $r_{\rm yr}$ & $r_{\rm yb}$ & $r_{\rm ym}$ & $r_{\rm gr}$ & $r_{\rm gb}$ & $r_{\rm gm}$ & $r_{\rm rb}$ & $r_{\rm rm}$ & $r_{\rm bm}$  \\
24 & $r_{\rm yg}$ & $r_{\rm ym}$ & $r_{\rm yr}$ & $r_{\rm yb}$ & $r_{\rm gm}$ & $r_{\rm gr}$ & $r_{\rm gb}$ & $r_{\rm mr}$ & $r_{\rm mb}$ & $r_{\rm rb}$  \\
\hline
\hline
\end{tabular}
\label{tab:sym}
\end{table}
  
\noindent
Similarly, multidimensional product kernels for 3 or 4-body
interaction energies can also be constructed from such 1D
kernels. Note that $p$-body interactions must be considered for all
permutations. For example, in the absence of symmetry, the CH$_4$
molecule has $^5C_4 = 5$ four body terms while including symmetry
there are $^5C_4 \times 4!  = 120$ four body terms. An explicit
example for all 2-, 3-, and 4-body terms for the case of CH$_2$O is
given in the supporting information.\\

\noindent
To determine all combinations of the 2-, 3-, and 4-body terms an
automated procedure is required that handles all possible symmetry
terms and also to eliminate redundancies. For this, an in-house
python\cite{python} code was written using the `itertools' module. The
software generates both, the required symmetrized form of the RKHS and
efficient fortran source code. A related strategy was followed
recently when constructing the fitting coefficients for PESs
represented as PIPs.\cite{qu18:151}\\

\subsection{Generation of the Reference Data Sets}
Although much higher levels of theory could in principle be used, in
particular for the smaller systems, the reference calculations in the
present work were carried out at the density functional theory (DFT)
level for convenience and illustration. All electronic structure
calculations were performed using the Orca 4.0\cite{orca} software
using the B3LYP functional\cite{bec93:5648,lee88:785} and
cc-pVDZ\cite{dun89:1007} basis set, similar to previous work on the
PIP-based PES for NMA.\cite{qu19:141101} `Very tight' SCF convergence
($10^{-9}$ hartree) criteria along with the largest grid (`grid7') for
the Lebedev integration were used in all calculations. The structures
of all molecules were optimized and harmonic frequencies were
determined. Then, reference structures were sampled using an in-house
written code as described in Ref. \citen{smi17:3192} at different
temperatures (20 to 2500 K) by distorting the equilibrium structures
and randomly displacing the atoms along the normal modes. For each of
the systems, energies and gradients were calculated for 4000 to 10000
reference structures. From this reference data, $N_{\rm train} = 1600$
to 2500 structures were used for constructing the RKHS (see Table
\ref{tab:qualitypes}) and $N_{\rm test} = 800$ to 1000 structures,
randomly drawn from the remaining data, were used for testing. Here,
it is worth to be mentioned that all structures with energies larger
than 4 eV with respect to the global minimum were excluded from the
reference and the test set. \\

\section{Results}
\subsection{Quality and Extrapolation of the PESs}
\begin{figure}[h!]
\includegraphics[scale=0.45]{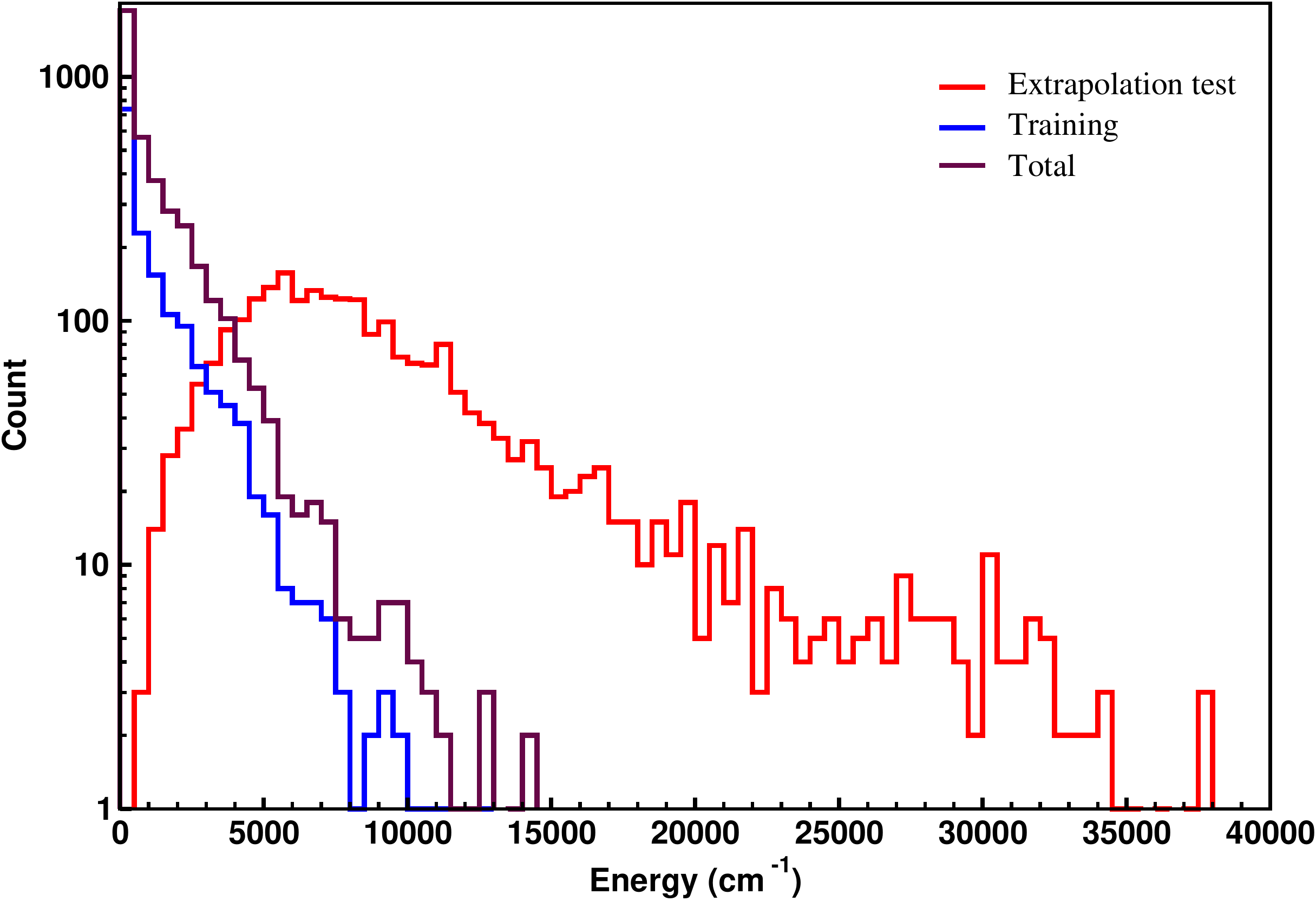}
\caption{Distribution of the reference and extrapolation data set for
  CH$_2$O. The distribution of the total data set (4001 points, brown)
  along with 1600 reference energies (blue lines) and 2500
  extrapolation energies (red lines). The counts are given on a
  logarithmic scale.}
\label{fig:fig2}
\end{figure}

First the quality of the resulting potential energy surfaces is
discussed. Unless otherwise stated, all RKHS-PESs were constructed
from using energies and gradients. As an example the data set
generated and used in constructing the multidimensional PESs for
formaldehyde is reported in Figure \ref{fig:fig2}. It shows the total
data set (brown), the reference set (blue), and the extrapolation set
(red) which extends to considerably higher energies. This last data
set is used to assess the extrapolation capabilities of the RKHS-based
PESs for structures (sampled at 5000 K), potentially far outside the
configurations used for generating the RKHS representation. One of the
potential shortcomings of certain machine learning approaches for
inter- and intramolecular PESs is their limitation as valid {\it
  interpolators} but not to extrapolate well beyond the structures
used to generate the model.\\

\begin{figure}[h!]
\centering \includegraphics[scale=1.5]{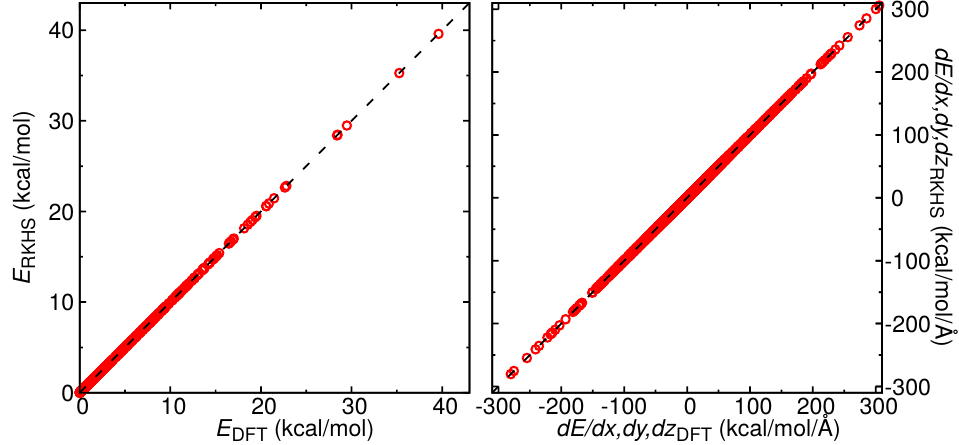}
\caption{Correlation between the energies (right) and gradients (left)
  for CH$_2$O molecule obtained from DFT calculations and predicted by
  the RKHS PES for 800 test data set. The RMSE and MAE for the PESs of
  all molecules are reported in Table \ref{tab:qualitypes}.}
\label{fig:fig3}
\end{figure}

\noindent
The performance of the RKHS-based PES for the test set is illustrated
in Figure \ref{fig:fig3}. Both, energies and forces are very
accurately described as the RMSE and MAE of 0.0003 kcal/mol and 0.0002
kcal/mol for energies and 0.004 kcal/mol/\AA\/ and 0.002
kcal/mol/\AA\/ for forces (gradients) demonstrate. For the coefficient
of determination, $R^2$, one finds $1-R^2 = 4 \times 10^{-9}$ and
$1-R^2 = 2 \times 10^{-8}$ for energies and forces, respectively, see
Table \ref{tab:qualitypes}.\\

\begin{table}[ht!]
\centering
\resizebox{\columnwidth}{!}{%
  \caption{Molecules, their sizes ($N_{\rm atom}\equiv n$), and the
    number of training N$_{\rm train}$ and test $N_{\rm test}$
    structures used. For each molecule the root mean squared error
    (RMSE), mean absolute error (MAE) for energies (kcal/mol) and
    forces (kcal/mol/\AA\/) and Pearson correlation coefficient
    calculated for $N_{\rm test}$ test data is given.}
  \label{tab:qualitypes}
    \begin{tabular}{l|ccc|ccc|ccc}
\hline
\hline
Molecule & $N_{\rm atom}$&$N_{\rm train}$ & $N_{\rm test}$ &\ \ \ RMSE \ \ \ & \ \ \ MAE \ \ \ & \ \ $1-R^2$ \ \ &\ \ \ RMSE \ \ \ & \ \ \ MAE \ \ \ & \ \ $1-R^2$ \ \ \\
\hline
&&&&\multicolumn{3}{c}{Energy}&\multicolumn{3}{|c}{Force}\\
CH$_2$O &4&1600&800& 0.0003&0.0002 & 4$\times 10^{-9}$& 0.0044 & 0.0021 & 2$\times 10^{-8}$ \\
CH$_4$ &5& 2400 &1000&0.0018& 0.0013& 9$\times 10^{-8}$&0.0098& 0.0048& 5$\times 10^{-7}$ \\
HCOOH &5& 2400 & 1000&0.0015 & 0.0007 & 2$\times 10^{-7}$& 0.0161 & 0.0069 & 2$\times 10^{-6}$ \\
CH$_3$OH &6& 2400 &1000& 0.0205& 0.0102& 5$\times 10^{-6}$ & 0.1064&0.0550& 6$\times 10^{-5}$\\
CH$_3$CHO &7&2400&1000& 0.0246&0.0124& 4$\times 10^{-6}$ &0.1067&0.0580 & 8$\times 10^{-5}$ \\
CH$_3$NO$_2$ &7&2500&1000 &0.0181& 0.0092& 1$\times 10^{-5}$ &0.0974&0.0525 & 9$\times 10^{-5}$ \\
CH$_3$COOH &8&2500&1000&0.0188&0.0093 & 6$\times 10^{-7}$&0.0919& 0.0483& 5$\times 10^{-5}$\\
CH$_3$CONH$_2$ &9&2500&1000 &0.0431&0.0132 & 2$\times 10^{-6}$&0.1190 &0.0571& 5$\times 10^{-5}$ \\
CH$_3$COCH$_3$ &10&2500&1000 &0.1019&0.0659 & 2$\times 10^{-5}$&0.3067 &0.2002& 3$\times 10^{-4}$\\
\hline
\hline
    \end{tabular}
    }
\end{table}

\begin{figure}[h!]
\centering
\includegraphics[scale=1.5]{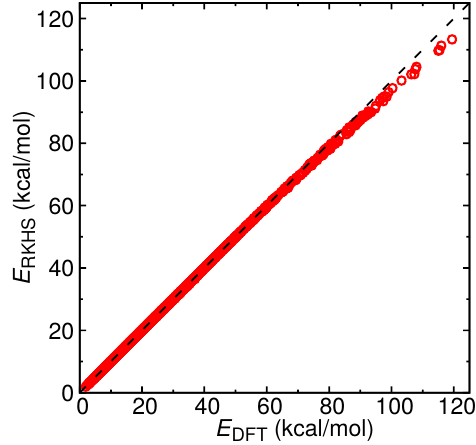}
\caption{Performance on the 2500 structures for CH$_2$O from the
  extrapolation data set (red line in Figure \ref{fig:fig2}), sampled
  at 5000 K. Correlation between the energies obtained from DFT
  calculations and predicted by the RKHS PES trained on energies and
  gradients for 1600 structures. The RKHS prediction has an RMSE of
  0.532 kcal/mol, MAE of 0.114 kcal/mol with $R^2 = 0.99913$).}
\label{fig:fig4}
\end{figure}

\noindent
Although the performance on the test data is very favourable, an even
more important aspect of molecular PESs in particular when used in
atomistic simulation is their validity and quality for structures far
away from those they were trained on. This is required for stable and
meaningful MD simulations. The extrapolation capability is
demonstrated in Figure \ref{fig:fig4} which demonstrates that the RKHS
PES for CH$_2$O remains accurate for energies three times higher than
for the energies in the reference and test set. Up to energies $\sim
100$ kcal/mol above the global minimum the RMSE is better than 0.5
kcal/mol which allows reliable MD simulations even at high
temperatures.\\

\noindent
The supporting information provides similar information for the CH$_4$
molecule, i.e. the energy distribution for all energies, those used
for constructing the RKHS-PES and those used for testing (see Figure
S2 and the validation of the RKHS-PES as the
correlation of energies and gradients between the reference
calculations and the evaluation of the RKHS-PES (Figure
S3). Very accurate predictions can also been achieved
in this case.\\

\begin{figure}
\centering
\includegraphics[scale=1.5]{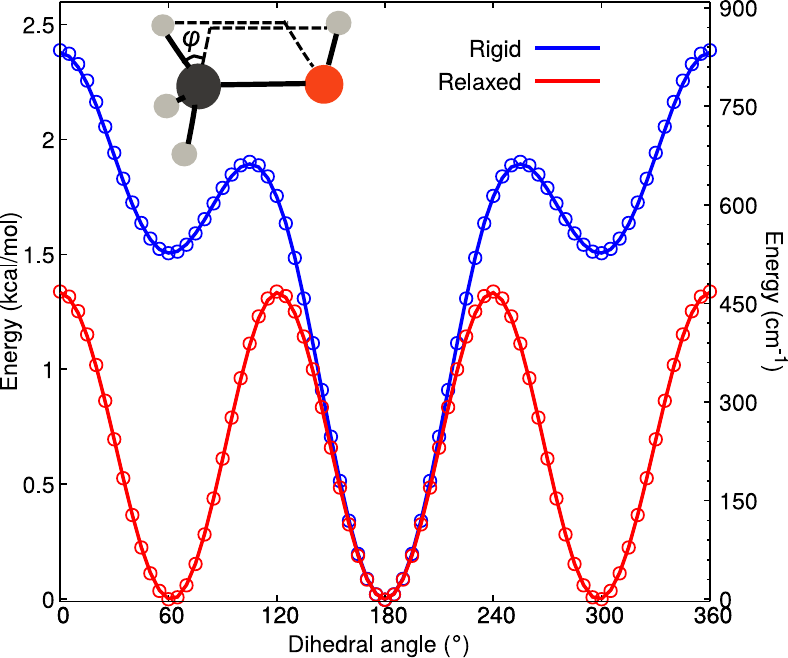}
\caption{Potential energies obtained from DFT calculations (open
  circles) and RKHS PES (solid lines) as a function of the H-C-O-H
  dihedral angle in CH$_3$OH. Blue line shows energies for a rigid
  scan changing only one H-C-O-H dihedral angle and the red line shows
  energies for a relaxed scan where the molecule is optimized for each
  value of the H-C-O-H angle. The definition of the dihedral angle is
  shown at top left; filled circles represent different atoms, gray
  black and red color represent the H, C and O atoms, respectively.}
\label{fig:figch3oh}
\end{figure}

\noindent
A typical cut through the global potential energy surface is afforded
by considering 1-dimensional energy functions along particular
internal degrees of freedom. One degree of freedom that is
particularly challenging in empirical energy function (``force
field'') development are dihedral torsions. Figure \ref{fig:figch3oh}
reports the potential energy profiles along the H-C-O-H torsion in
CH$_3$OH for a rigid and a relaxed scan. In a rigid scan potential
energies are calculated for different values of the H-C-O-H dihedral
angle while keeping all other degrees of freedom frozen at the
equilibrium geometry. Conversely, in a relaxed scan the structure of
the molecule is optimized for a given value of the H-C-O-H dihedral
angle. Both scans from the RKHS PES accurately reproduce the reference
B3LYP data. The symmetry of the molecule (i.e. permutations among the
methyl hydrogens) is also preserved in the RKHS PES.\\

\noindent
To quantify the advantage of the ``energy+gradient'' based RKHS method
over the ``energy-only'' data set (where only energies are used as an
input to obtain the coefficients, see Eq. \ref{rkhs}) energy and force
learning curves on the test data sets are calculated for CH$_4$. The
``learning curves'' for the RMSE (red lines) and MAE (blue lines) for
both energies and forces, using ``energy only'' (dashed) and
``energy+gradient'' (solid), are shown in Figure
S4. When using ``energy only'' (dashed curves), both,
energies (left panel) and forces (right panel) continuously improve as
the size of the training set increases and further improvements appear
to be possible beyond $6 \times 10^{-4}$ kcal/mol for energies and $6
\times 10^{-3}$ kcal/mol/\AA\/ for the largest training set ($N_{\rm
  train} = 9600$). However, for the forces the ``energy+gradient''
approach reaches similar accuracy as the ``energy only'' RKHS using
1/6 of the data (i.e. $N_{\rm train}^{\rm energy+gradient} = 1600$ vs.
$N_{\rm train}^{\rm energy} = 9600$). Hence, including gradient
information explicitly in the RKHS, see Eq. \ref{rkhs+f}, reduces the
number of coefficients which also speeds up the RKHS evaluation. The
energy learning curves from using ``energy+gradient'' in constructing
the RKHS-PESs appear to saturate with ($N_{\rm train} = 3200$) at
similar values for RMSD and MAE. This is because the weights of the
gradients are $3n$ times larger than those for the energies, where $n$ is
the total number of atoms of the molecule.\\

\subsection{Quality of Normal Mode Frequencies from RKHS-PESs}
Normal mode frequencies are useful computational observables to
compare the performance of fitted PESs with the reference calculations
they are based on.\cite{qu19:141101} Harmonic frequencies were
calculated for the molecules using the ASE package\cite{Larsen:2017}
by linking the RKHS PESs as an external energy calculator. Table
\ref{tab:ch2ofreq} compares the normal mode frequencies from the
B3LYP/cc-pVDZ calculations with those from the RKHS-represented PESs
for CH$_2$O, HCOOH, and CH$_4$. Besides the remarkable accuracy
(difference $< 1$ cm$^{-1}$ for every mode) with which the
kernel-represented PESs are capable of describing the reference
calculation for all examples considered, maintaining the correct
symmetry and degeneracy in the case of CH$_4$ is most notable. In
particular, the RKHS PES exactly (for the HCH bend) or very closely
(for the CH stretch) maintains the two triply degenerate modes at 1309
cm$^{-1}$ and 3146 cm$^{-1}$, respectively, as it should be. This also
underlines the correct implementation of permutational invariance in
the formulation.{}\\

\begin{table}[ht!]
\caption{Harmonic frequencies (in cm$^{-1}$ and rounded to full
  wavenumbers) and zero point energies (in eV) for CH$_2$O, HCOOH and
  CH$_4$ computed using the reference B3LYP/cc-pVDZ calculations
  (Ref.) and calculated from their RKHS-PES (RKHS). The RKHS-PESs were
  trained on energies and gradients. The RMSD between reference values
  and those from the RKHS PESs is well below 1 cm$^{-1}$.}
    \begin{tabular}{c|cc|cc|cc}
\hline
\hline
& \multicolumn{2}{c|}{CH$_2$O}&\multicolumn{2}{c}{HCOOH}&\multicolumn{2}{|c}{CH$_4$}\\
mode & Ref. & \ \ \ RKHS \ \ \ & Ref. & \ \ \ RKHS \ \ \ & Ref. & \ \ \ RKHS \ \ \ \\
\hline
  1 & 1186 &  1186 & 627 &  627 & 1309 &  1309 \\
  2 & 1252 &  1252 & 700 &  701 & 1309 &  1309\\
  3 & 1514 &  1514 & 1046 & 1046& 1309 &  1309\\
  4 & 1831 &  1831 & 1138 & 1137& 1530 &  1529\\
  5 & 2862 &  2862 & 1311 & 1311& 1530 &  1530\\
  6 & 2914 &  2914 & 1394 & 1393& 3025 &  3025\\
  7 &           &           & 1843 & 1843 & 3146 &  3145\\
  8 &           &           & 3031 & 3031 & 3146 &  3146\\
  9 &           &           & 3676 & 3677 & 3146 &  3146\\
  \hline 
  ZPE &  0.717 & 0.717 & 0.916& 0.917 & 1.206 &1.206 \\
\hline
\hline
\end{tabular}
\label{tab:ch2ofreq}
\end{table}

\begin{figure}[h!]
\centering
\includegraphics[scale=1.7]{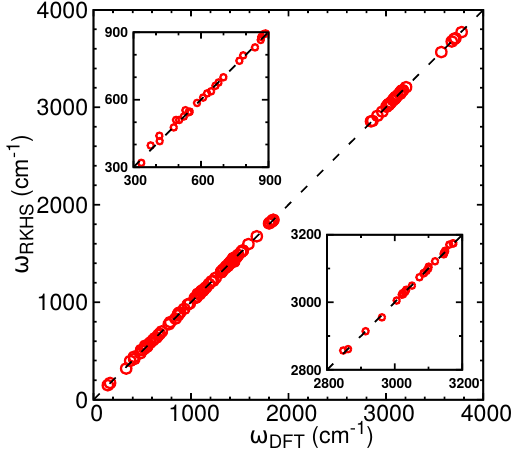}
\caption{Correlation between the harmonic frequencies for all the
  systems considered obtained from DFT calculations and RKHS PESs with
  an RMSE of 6.7 cm$^{-1}$ and MAE of 4.1 cm$^{-1}$, and $R^2 =
  0.99995$. The insets show magnifications of the low- and
  high-frequency vibrations. All RKHS PESs are based on
  energy+gradients.}
\label{fig:fig5}
\end{figure}

\noindent
A broader overview of all harmonic frequencies for all compounds in
Table \ref{tab:qualitypes} is shown in Figure \ref{fig:fig5}. These
normal mode frequencies are from the RKHS-PESs trained on energies and
gradients. For the 124 normal mode frequencies the overall MAE between
reference calculations and frequencies determined on the RKHS-PESs is
4.1 cm$^{-1}$ with $R^2 = 0.99995$. This is consistent with the high
accuracy of the energies and forces reported in Table
\ref{tab:qualitypes}. Here it is worth to be mentioned that for larger
molecules low frequency ($ < 200 $ cm$^{-1}$) modes contribute most to
the error. This is consistent with recent work using PIPs for a
full-dimensional PES for N-methyl acetamide for which some of the
low-frequency modes differ up to $\sim$ 30
cm$^{-1}$.\cite{qu19:141101} It should be emphasised that such
accuracy is independent of the quality of the electronic structure
method used for the reference calculations. In other words, if
energies and forces are available at a considerably higher level of
theory (e.g. CCSD(T) with a large basis set) the same performance in
reproducing such reference data as that reported here is expected
which provides a very high accuracy but computationally efficient
energy function with analytical gradients.\\

\begin{figure}[h!]
\centering \includegraphics[scale=1.3]{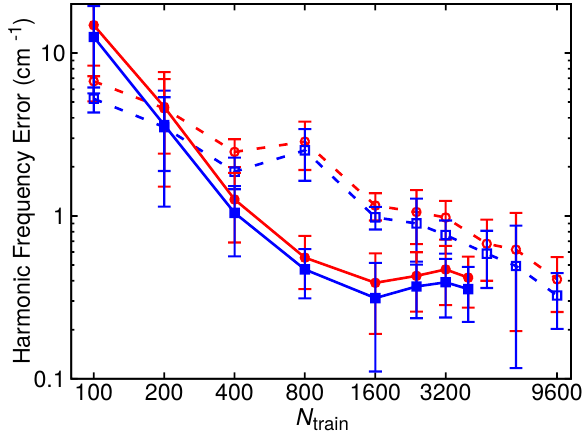}
\caption{Root mean squared difference for harmonic frequencies for
  CH$_4$ from using ``energy only'' (dashed lines and open symbols)
  and ``energy+gradient'' (solid lines and filled symbols) training. For
  a given number of training data each model is trained for five times
  for random data set. Average values and standard deviations (error
  bars) of the RMSE and MAE are shown as red and blue, respectively.}
\label{fig:fig9}
\end{figure}

\noindent
Another property of interest concerns the change (ideally
``improvement'') of an observable (here normal modes) as the number of
training data $N_{\rm train}$ increases. This is reported in Figure
\ref{fig:fig9} for RKHS-PESs trained on ``energies only'' and
``energies + gradients''. When energies only are used for training the
RKHS PES for CH$_4$ an average error better than 1 cm$^{-1}$ requires
$N_{\rm train} \sim 3200$ training data whereas including energies and
gradients in generating the RKHS-PES already achieves this with
$N_{\rm train} \sim 400$. This should be compared with the findings
for the learning curves in Figure S3 that report a
similar performance for ``energy only'' and ``energy+gradients'' for
$N_{\rm ref} = 9600$ and $N_{\rm ref} = 1600$, respectively. This is
attributed to the additional information the gradients provide about
the local curvature around every structure for which an energy is
available. Furthermore, the curves in Figure \ref{fig:fig9} behave
very differently for ``energy only'' and ``energy+gradients'' used in
constructing the RKHS-PES. Whereas the PES trained on ``energies
only'' appears to have two slopes (up to $N_{\rm train} \sim 400$ and
beyond $N_{\rm train} > 800$ with a local maximum deviation at $N_{\rm
  train} \sim 800$), normal modes determined on the
``energy+gradients'' trained PESs continuously improve until $N_{\rm
  train} \sim 1600$ to an average error of 0.2 cm$^{-1}$ after which
they level off within the fluctuation bars. Probably this is the
maximum accuracy that can be achieved for harmonic frequencies. Again
it is to be mentioned that the Hessian is calculated numerically in
ASE.\\

\section{Discussion and Conclusions}
The present work introduces an extension of RKHS-based
PESs\cite{ho96:2584} to polyatomic molecules. Combining energy and
force information to construct tensor-product based kernels up to
4-body interactions is shown to yield highly accurate PESs for
molecules ranging from formaldehyde to acetone. Using ``energy +
gradients'' for constructing the RKHS-PES requires between a factor of 6
to 10 less reference data than working with ``energy only''. The
RKHS-PESs are very accurate and extrapolate well to structures with
considerably higher energies, see Figure \ref{fig:fig4}. This is not
guaranteed for NN-learned PESs as recent work on
acetaldehyde\cite{MM.atmos:2020} with the PhysNet\cite{unk19:3678}
NN-architecture has shown. Unless structures at the highest energies
are included, many of the MD trajectories become invalid as the
energies and forces generated from the NN are inconsistent with the
true energies and forces compared with the reference electronic
structure calculations.\\

\noindent
The harmonic modes computed from the RKHS PES and from the reference
electronic structure calculations (here B3LYP/cc-pVDZ) are within 1
cm$^{-1}$ for small molecules and within 5 cm$^{-1}$ for larger
molecules except for low frequency modes ($<200$ cm$^{-1}$). Similar
observation were also made for {\it cis-} and {\it trans}-NMA using
PIP-based PESs.\cite{qu19:141101,nan19:084306} Such performance
naturally extends to reference data computed at a much higher level of
theory. Hence, for systems with up to 10 atoms considered here the
only limitation will be the computing time required for generating the
training and test data set.\\

\begin{figure}
\centering \includegraphics[scale=1.6]{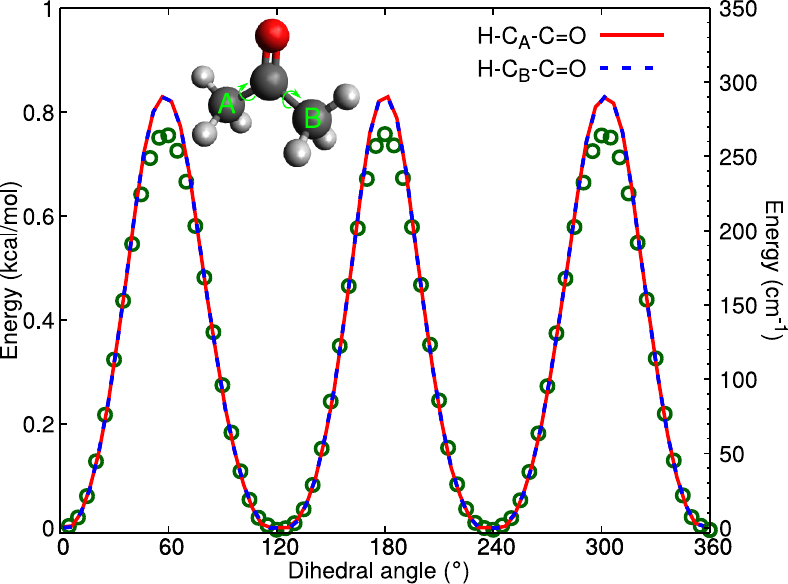}
\caption{Potential energies obtained from B3LYP/cc-pVDZ calculations
  (green open circles) and RKHS PES (solid red and dashed blue lines)
  as a function of H-C-C-O dihedral angles in CH$_3$COCH$_3$
  (acetone). A relaxed scan is performed for both the dihedral angles
  where the molecule is optimized for each points.}
\label{fig:figacetone}
\end{figure}

\noindent
To achieve an agreement between reference data and that from the
representation (here RKHS) for arbitrary configurations or even
low-dimensional projections (e.g. a torsional potential) for bonded
terms is extremely challenging for empirical force fields.  As an
example, earlier versions of the CHARMM force field\cite{charmm22} had
to be empirically corrected by introducing the CMAP
correction\cite{brooks:2004} to account for deficiencies in the
dihedral potentials. Because the number of dihedral terms is large and
primarily responsible for secondary and tertiary structural changes in
peptides and proteins, specifically improving these contributions to
empirical force fields appears to be a useful possibility. It is also
worth to point out that the RKHS PES is permutationally invariant for
the equivalent methyl H atoms which is also seen in Figure
\ref{fig:figch3oh}. These findings also extend to larger molecules as
demonstrated for dihedral scans for acetone as reported in Figure
\ref{fig:figacetone}. The relaxed scan from the reference
B3LYP/cc-pVDZ calculations and the RKHS PES agree very well except
around the top of the barrier where they differ by $\sim 25$
cm$^{-1}$. Both the methyl group and also the methyl hydrogens in each
group preserved their symmetry in the RKHS PES. \\

\begin{figure}[h!]
\centering \includegraphics[scale=1.4]{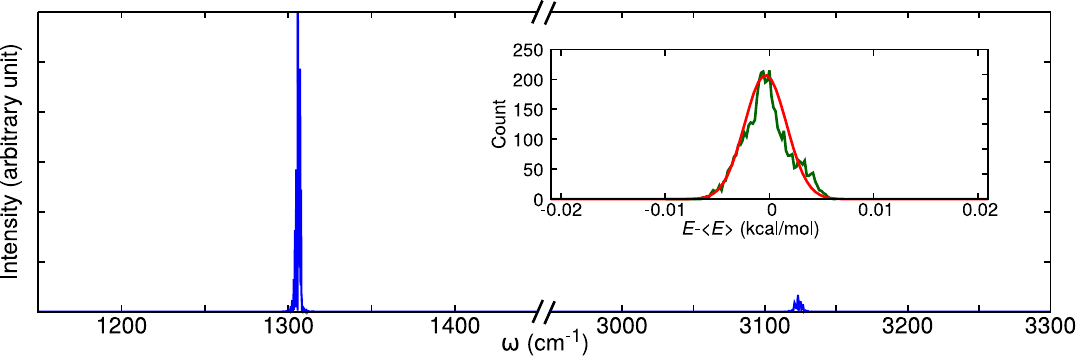}
\caption{IR spectrum for CH$_4$ obtained from the dipole moment
  autocorrelation function and subsequent fast Fourier
  transformation. The molecular dipole moment was computed by using
  Mulliken point charges from DFT calculations for the equilibrium
  structure. The infrared active modes (triply degenerate HCH bend and
  triply degenerate CH stret modes) are at 1306 cm$^{-1}$ and 3123
  cm$^{-1}$, respectively. As required, the totally symmetric,
  infrared inactive CH stretch mode at 3025 cm$^{-1}$, see Table
  \ref{tab:ch2ofreq}, does not appear in the infrared spectrum. The
  inset shows the distribution of the total energy fluctuation around
  its $\langle E \rangle$ (green line) in the MD simulations with a
  superimposed Gaussian function (red line).}
\label{fig:fig10}
\end{figure}

\noindent
Another future application of the methods discussed here are molecular
dynamics simulations of small molecules on global, anharmonic and
fully coupled RKHS PESs. As an example, the infrared spectrum for
CH$_4$ in the gas phase is reported in Figure \ref{fig:fig10}. This
simulation was carried out with a suitably modified version of the
CHARMM molecular simulation program\cite{Brooks.charmm:2009} to use
energies and forces from the RKHS-PES. The PES trained on 2400
structures using both energies and gradients was used. The time step
in this simulation was 0.1 fs and the simulation temperature was 300
K. First, the system is heated to the simulation temperature,
equilibrated for 7 ps and then an equilibrium $NVE$ simulation was
carried out for 250 ps. Total energy is conserved to within 0.015
kcal/mol, see inset of Figure \ref{fig:fig10}, which underlines that
the forces in the RKHS are correctly implemented.\\

\noindent
Finally, the possibility to extend the methodology introduced here to
intermolecular interactions is mentioned. The present work was
concerned with the ``bonded interactions'' when comparing with
empirical force field technology.\cite{charmm36ff,amberff,oplsff}
However, for condensed phase simulations, nonbonded interactions
between, e.g., a solute and the surrounding solvent need to be
determined and available as well. One future possibility is to combine
the accurate RKHS-PESs discussed here with accurate multipolar
electrostatic models (possibly augmented by
polarization).\cite{MM.mtp:2012,MM.mtp:2013} Alternatively, developing
an RKHS-based fragment approach can be envisaged to treat molecular
dimers and trimers. \\

\noindent
In conclusion, the RKHS technique which has already been found to be
highly beneficial for the study of reactive
processes\cite{kon18:094305,MM.n2ar:2019,kon20n2o,san20:3927} and
spectroscopic studies\cite{MM.n3:2019,kon19:24976,kon20n3p} has been
considerably extended to treat the intramolecular degrees of freedom
for molecules with up to 10 atoms. Together with further developments
this approach is expected to provide a way towards quantitative gas-
and condensed-phase simulations.\\

\section*{Acknowledgment}
The authors acknowledge financial support from the Swiss National
Science Foundation (NCCR-MUST and Grant No. 200021-7117810), the
AFOSR, and the University of Basel.

\clearpage
\providecommand{\latin}[1]{#1}
\makeatletter
\providecommand{\doi}
  {\begingroup\let\do\@makeother\dospecials
  \catcode`\{=1 \catcode`\}=2 \doi@aux}
\providecommand{\doi@aux}[1]{\endgroup\texttt{#1}}
\makeatother
\providecommand*\mcitethebibliography{\thebibliography}
\csname @ifundefined\endcsname{endmcitethebibliography}
  {\let\endmcitethebibliography\endthebibliography}{}

\end{document}


\section{Symmetrized RKHS for CH$_2$O}
Here the explicit expressions for non-symmetrized and symmetrized
preserving permutational invariance for 2-, 3-, and 4-body terms are
given explicitly for formaldehyde.

\subsection{2-, 3- and 4-body Terms Without Symmetry}
\begin{figure}[h!]
\centering \includegraphics[scale=1.4]{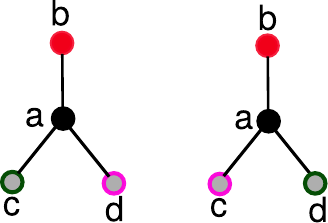}
\caption{The two permutationally invariant configurations of CH$_2$O
  molecule, configuration 1 (left) and configuration 2 (right). Atoms
  are represented by color, black (b) is for the carbon atom, red (r)
  is for the oxygen atom, and gray with green border (g) and gray with
  magenta border (m) for the hydrogen atoms. Positions of the atoms
  are denoted by `a', `b', `c', and `d'.}
\label{fig:sifig1}
\end{figure}

Formaldehyde has two permutationally invariant hydrogen atoms which
leads to two equivalent configurations for any structure, see Figure
\ref{fig:sifig1}. Table \ref{tab:sitab1} shows the permutationally
invariant interatomic distances for the two equivalent configurations,
configuration 1 and configuration 2. Without symmetry, there are six
different interatomic distances, i.e. $r_{\rm ab}$, $r_{\rm ac}$,
$r_{\rm ad}$, $r_{\rm bc}$, $r_{\rm bd}$ and $r_{\rm cd}$. They form
six 1D kernel basis functions ($6\times1^2$) for the total kernel
polynomial, i.e., $k(r_{\rm ab},r_{\rm ab}'), k(r_{\rm ac},r_{\rm
  ac}'), \cdots, k(r_{\rm cd},r_{\rm cd}')$.\\

\begin{table}[ht!]
\caption{Symmetrization order of interatomic distances for two
  equivalent CH$_2$O configurations. Interatomic distances between two
  different atoms/positions are $r_{ij}$ = $r_{ji}$. Atom positions
  and color indices are defined in Figure \ref{fig:sifig1}.}
    \begin{tabular}{l|c|cc|cc|c}
\hline
\hline
Configuration & $r_{\rm ab}$ & $r_{\rm ac}$ & $r_{\rm ad}$ & $r_{\rm bc}$ & $r_{\rm bd}$ & $r_{\rm cd}$ \\
\hline
1  & $r_{\rm br}$ & $r_{\rm bg}$ & $r_{\rm bm}$ & $r_{\rm rg}$ & $r_{\rm rm}$ & $r_{\rm gm}$ \\
2  & $r_{\rm br}$ & $r_{\rm bm}$ & $r_{\rm bg}$ & $r_{\rm rm}$ & $r_{\rm rg}$ & $r_{\rm mg}$ \\
\hline
\hline
\end{tabular}
\label{tab:sitab1}
\end{table}

\noindent
In the following, the 1D kernels used for the 2-, 3-, and 4-body terms
are $k^{[3,5]}$, $k^{[3,1]}$, and $k^{[3,0]}$, see main text. Then the
unsymmetrized 2-body interaction energy of CH$_2$O can be expressed as
a sum of these 1D kernel functions,
\begin{multline*}
K_{\rm 2b}({\bf r}, {\bf r'}) = k^{[3,5]}(r_{\rm ab},r_{\rm
  ab}')+k^{[3,5]}(r_{\rm ac},r_{\rm ac}')+k^{[3,5]}(r_{\rm ad},r_{\rm
  ad}') +k^{[3,5]}(r_{\rm bc},r_{\rm bc}')+ \\k^{[3,5]}(r_{\rm
  bd},r_{\rm bd}')+ k^{[3,5]}(r_{\rm cd},r_{\rm cd}').
\end{multline*}
There are four ($^4C_3$) 3-body interactions in the CH$_2$O
molecule. Each 3-body interaction energy is a by product of three 1D
kernel functions. Without considering symmetry the 3-body interaction
energies are
\begin{multline*}
K_{\rm 3b}({\bf r}, {\bf r'}) = k^{[3,1]}(r_{\rm ab},r_{\rm
  ab}')\times k^{[3,1]}(r_{\rm ac},r_{\rm ac}')\times k^{[3,1]}(r_{\rm
  bc},r_{\rm bc}')\\ +k^{[3,1]}(r_{\rm ab},r_{\rm ab}')\times
k^{[3,1]}(r_{\rm ad},r_{\rm ad}')\times k^{[3,1]}(r_{\rm bd},r_{\rm
  bd}')\\ +k^{[3,1]}(r_{\rm ac},r_{\rm ac}')\times k^{[3,1]}(r_{\rm
  ad},r_{\rm ad}')\times k^{[3,1]}(r_{\rm cd},r_{\rm
  cd}')\\ +k^{[3,1]}(r_{\rm bc},r_{\rm bc}')\times k^{[3,1]}(r_{\rm
  bd},r_{\rm bd}')\times k^{[3,1]}(r_{\rm cd},r_{\rm cd}').
\end{multline*}
There is only one ($^4C_4$) 4-body interaction in the CH$_2$O
molecule. Each 4-body interaction energy can be written as a product
of six 1D kernel functions. Without considering symmetry the 4-body
interaction energies can be written as
\begin{multline*}
K_{\rm 4b}({\bf r}, {\bf r'}) = k^{[3,0]}(r_{\rm ab},r_{\rm
  ab}')\times k^{[3,0]}(r_{\rm ac},r_{\rm ac}')\times k^{[3,0]}(r_{\rm
  ad},r_{\rm ad}') \times k^{[3,0]}(r_{\rm bc},r_{\rm bc}')\times
\\k^{[3,0]}(r_{\rm bd},r_{\rm bd}')\times k^{[3,0]}(r_{\rm cd},r_{\rm
  cd}').
\end{multline*}

\subsection{2-, 3- and 4-body Terms With Symmetry}
If two fold symmetry is included, there are four types of interatomic
distances in the CH$_2$O molecule i.e. one CO, two CH, two OH and one
HH. They form 10 ($1^2$+$2^2$+$2^2$+$1^2$) 1D kernel basis functions
for the total kernel polynomial, and include $[k(r_{\rm br},r_{\rm
    br}')$, $k(r_{\rm bg},r_{\rm bg}')$, $k(r_{\rm bm},r_{\rm bm}')$,
  $k(r_{\rm rg},r_{\rm rg}')$, $k(r_{\rm rm},r_{\rm rm}')$, $k(r_{\rm
    gm},r_{\rm gm}')]$ for the CO, CH, OC, and HH distances of
configuration 1, and $[k(r_{\rm br},r_{\rm br}')$, $k(r_{\rm
    bg},r_{\rm bm}')$, $k(r_{\rm bm},r_{\rm bg}')$, $k(r_{\rm
    rg},r_{\rm rm}')$, $k(r_{\rm rm},r_{\rm rg}')$, $k(r_{\rm
    gm},r_{\rm mg}')]$ for configuration 2. It should be noted that
there are 12 kernel functions (see Table \ref{tab:sitab1}) of which
$k(r_{\rm br},r_{\rm br}')$ and $k(r_{\rm gm},r_{\rm gm}')$ appear
twice.\\

\noindent
The symmetrized 2-body interaction energy of CH$_2$O is then a sum of
these 12 1D kernel functions
\begin{multline*}
K_{\rm 2b}({\bf r}, {\bf r'}) = 2\times k^{[3,5]}(r_{\rm br},r_{\rm
  br}')+k^{[3,5]}(r_{\rm bg},r_{\rm bg}')+k^{[3,5]}(r_{\rm bm},r_{\rm
  bm}') +k^{[3,5]}(r_{\rm rg},r_{\rm rg}')+ \\ k^{[3,5]}(r_{\rm
  rm},r_{\rm rm}')+k^{[3,5]}(r_{\rm bg},r_{\rm bm}')+ k^{[3,5]}(r_{\rm
  bm},r_{\rm bg}')+k^{[3,5]}(r_{\rm rg},r_{\rm rm}')+ k^{[3,5]}(r_{\rm
  rm},r_{\rm rg}')+2\times k^{[3,5]}(r_{\rm gm},r_{\rm gm}')
\end{multline*}

\noindent
For each equivalent configuration there are four ($^4C_3$) 3-body
interactions in the CH$_2$O molecule. Considering symmetry the 3-body
interaction energy is
\begin{multline*}
K_{\rm 3b}({\bf r}, {\bf r'}) = k^{[3,1]}(r_{\rm br},r_{\rm
  br}')\times k^{[3,1]}(r_{\rm bg},r_{\rm bg}')\times k^{[3,1]}(r_{\rm
  rg},r_{\rm rg}')\\ +k^{[3,1]}(r_{\rm br},r_{\rm br}')\times
k^{[3,1]}(r_{\rm bg},r_{\rm bm}')\times k^{[3,1]}(r_{\rm rg},r_{\rm
  rm}')\\ +k^{[3,1]}(r_{\rm br},r_{\rm br}')\times k^{[3,1]}(r_{\rm
  bm},r_{\rm bg}')\times k^{[3,1]}(r_{\rm rm},r_{\rm
  rg}')\\ +k^{[3,1]}(r_{\rm br},r_{\rm br}')\times k^{[3,1]}(r_{\rm
  bm},r_{\rm bm}')\times k^{[3,1]}(r_{\rm rm},r_{\rm
  rm}')\\ +k^{[3,1]}(r_{\rm bg},r_{\rm bg}')\times k^{[3,1]}(r_{\rm
  bm},r_{\rm bm}')\times k^{[3,1]}(r_{\rm gm},r_{\rm
  gm}')\\ +k^{[3,1]}(r_{\rm bg},r_{\rm bm}')\times k^{[3,1]}(r_{\rm
  bm},r_{\rm bg}')\times k^{[3,1]}(r_{\rm gm},r_{\rm
  gm}')\\ +k^{[3,1]}(r_{\rm rg},r_{\rm rg}')\times k^{[3,1]}(r_{\rm
  rm},r_{\rm rm}')\times k^{[3,1]}(r_{\rm gm},r_{\rm
  gm}')\\ +k^{[3,1]}(r_{\rm rg},r_{\rm rm}')\times k^{[3,1]}(r_{\rm
  rm},r_{\rm rg}')\times k^{[3,1]}(r_{\rm gm},r_{\rm gm}')\\
\end{multline*}

\noindent
Finally, the two fold symmetry, 4-body interaction energies can be
written as
\begin{multline*}
K_{\rm 4b}({\bf r}, {\bf r'}) = k^{[3,0]}(r_{\rm br},r_{\rm
  br}')\times k^{[3,0]}(r_{\rm bg},r_{\rm bg}')\times k^{[3,0]}(r_{\rm
  bm},r_{\rm bm}') \times k^{[3,0]}(r_{\rm rg},r_{\rm rg}')\times
\\k^{[3,0]}(r_{\rm rm},r_{\rm rm}')\times k^{[3,0]}(r_{\rm gm},r_{\rm
  gm}')\\ +k^{[3,0]}(r_{\rm br},r_{\rm br}')\times k^{[3,0]}(r_{\rm
  bg},r_{\rm bm}')\times k^{[3,0]}(r_{\rm bm},r_{\rm bg}') \times
k^{[3,0]}(r_{\rm rg},r_{\rm rm}')\times \\k^{[3,0]}(r_{\rm rm},r_{\rm
  rg}')\times k^{[3,0]}(r_{\rm gm},r_{\rm gm}')
\end{multline*}

\section{CH$_4$ RKHS PES}
\begin{figure}[h!]
\centering
\includegraphics[scale=0.6]{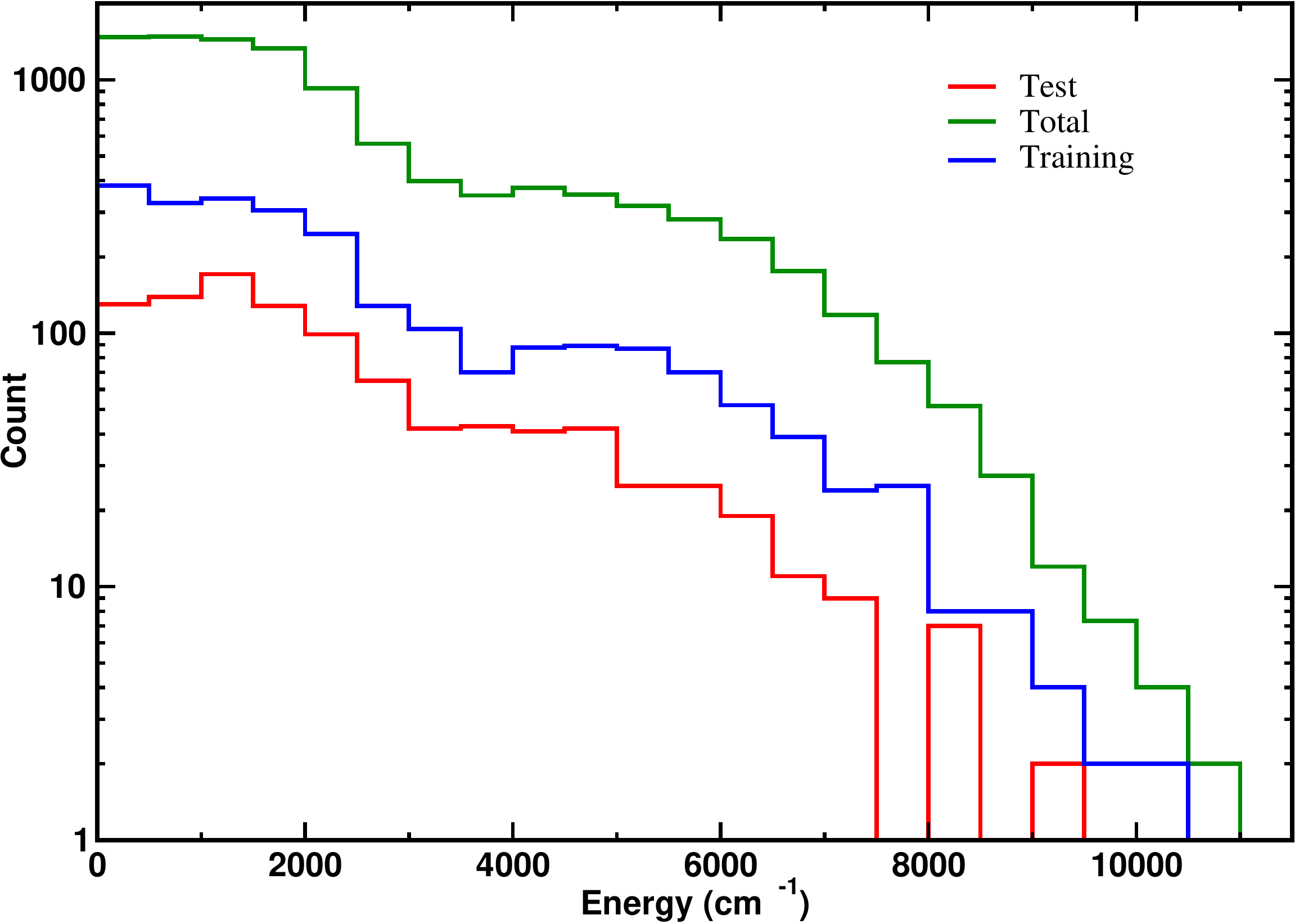}
\caption{Distribution of the reference data set for
  CH$_4$. Distribution of all 10000 reference energies (green) along
  with 2400 training energies (blue) and 1000 test energies (red).}
\label{sifig:ch4dist}
\end{figure}

\begin{figure}[h!]
\centering \includegraphics[scale=1.5]{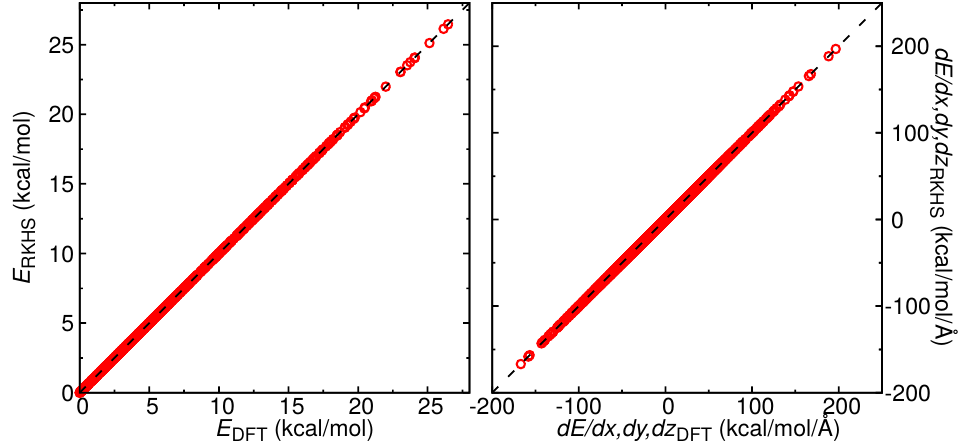}
\caption{Correlation between reference (DFT) and prediction by the
  RKHS-PES for energies (left) and gradients (right) for CH$_4$ for
  $N_{\rm test} = 1000$. The RMSE, MAE and $(1-R^2)$ for the energies
  are 0.0018, 0.0013 kcal/mol and $9\times 10^{-8}$, respectively. For
  the gradients the RMSE, MAE and $(1-R^2)$ are 0.0098, 0.0048
  kcal/mol/\AA\ and $5\times 10^{-7}$, respectively.}
\label{sifig:ch4corr}
\end{figure}

\begin{figure}[h!]
\centering \includegraphics[scale=1.3]{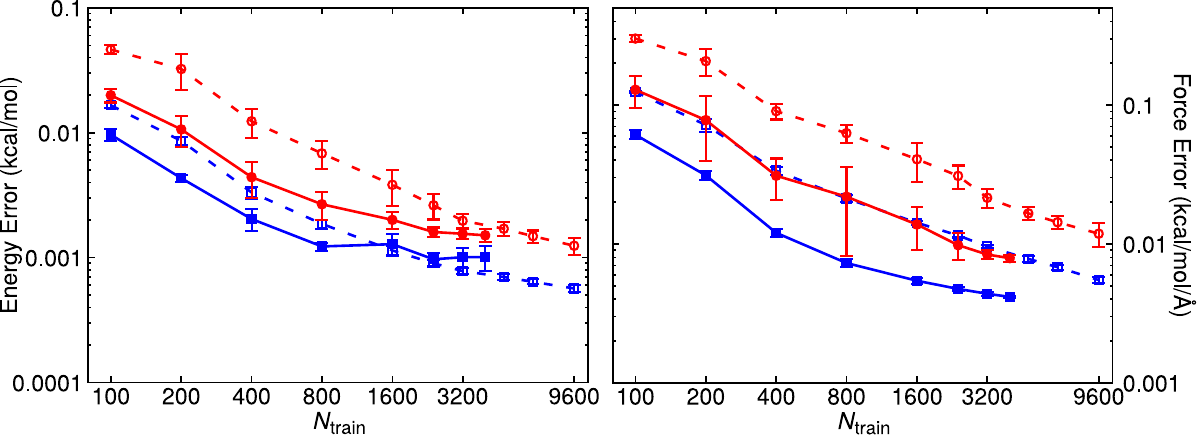}
\caption{Energy (left) and force (right) learning curve for CH$_4$ for
  ``energy only'' (dashed lines and open symbols) and ``energy+force''
  (solid lines and filled symbols). For a given number $N_{\rm ref}$
  each model is generated five times using randomly drawn reference
  data. Average values and standard deviations (error bars) of the
  RMSE (red) and MAE (blue) are shown, respectively.}
\label{sifig:learn}
\end{figure}